\documentclass[conference]{IEEEtran}
\IEEEoverridecommandlockouts
\usepackage{cite}
\usepackage{amsmath,amssymb,amsfonts}
\usepackage{algorithmic}
\usepackage{graphicx}
\usepackage{textcomp}
\usepackage{xcolor}
\usepackage{multirow}
\usepackage{comment, amsfonts}
\usepackage{tabularx}
\usepackage{subcaption}
\usepackage{makecell}
\usepackage[normalem]{ulem}

\usepackage{kotex}
\newcommand{\indicator}[1]{\mathbb{I}\left[#1\right]}

\def\BibTeX{{\rm B\kern-.05em{\sc i\kern-.025em b}\kern-.08em
    T\kern-.1667em\lower.7ex\hbox{E}\kern-.125emX}}
\begin{document}

\title{Token-based Attractors and Cross-attention in\\Spoof Diarization}

\author{Kyo-Won Koo$^{1,*}$, Chan-yeong Lim$^{1,*}$, Jee-weon Jung$^{2\ddag}$\thanks{$\ddag$: Currently at Apple.}, Hye-jin Shim$^{2}$, Ha-Jin Yu$^{1,\dagger}$\\
$^{1}$University of Seoul, Seoul, Korea\\
$^{2}$Carnegie Mellon University, Pittsburgh, PA, USA\\
kkwr0504@uos.ac.kr, cksdud585@naver.com, jeeweonj@ieee.org,\\
shimhz6.6@gmail.com, hjyu@uos.ac.kr\thanks{$^{*}$Equal contribution.\;$\dagger$Corresponding author.}\thanks{This research was supported by the National Research Foundation of Korea (NRF) grant funded by the korea government (MSIT) (NO. RS-2022-NR068754).}}

\maketitle

\begin{abstract}
Spoof diarization identifies ``what spoofed when" in a given speech by temporally locating spoofed regions and determining their manipulation techniques. 
As a first step toward this task, prior work proposed a two-branch model for localization and spoof type clustering, which laid the foundation for spoof diarization.
However, its simple structure limits the ability to capture complex spoofing patterns and lacks explicit reference points for distinguishing between bona fide and various spoofing types. 
To address these limitations, our approach introduces learnable tokens where each token represents acoustic features of bona fide and spoofed speech.
These attractors interact with frame-level embeddings to extract discriminative representations, improving separation between genuine and generated speech.
Vast experiments on PartialSpoof dataset consistently demonstrate that our approach outperforms existing methods in bona fide detection and spoofing method clustering.

\end{abstract}

\begin{IEEEkeywords}
Spoof diarization, Speaker diarization, Attractor tokens, Cross-attention
\end{IEEEkeywords}

\section{Introduction}
\noindent
Recent advances in speech synthesis and voice conversion technologies have enabled the generation of manipulated speech that closely resembles real human voices~\cite{triantafyllopoulos2023overview, li2023styletts, tan2024naturalspeech}.
While these technologies offer valuable applications, they also introduce security threats when misused, such as spoofing attacks that manipulate a person's voice to bypass automatic speaker verification systems.
To address such risks, the ASVspoof challenge series has fostered research in audio deepfake detection, focusing on distinguishing between bona fide and spoofed speech~\cite{wu2015asvspoof, kinnunen2017asvspoof, yamagishi2019asvspoof, delgado2021asvspoof, wang2025asvspoof}.

Early approaches primarily trained deep neural network-based countermeasures (CMs) to perform binary classification at the utterance level by learning global representations of fully spoofed speech~\cite{jung2022aasist, tak2021end, shin2024hm, wu24b_interspeech, STCA,CMFF}.
However, in real-world settings, attacks can also manipulate only a few words or phonemes, rather than synthesizing the entire utterance.
This form of attack, known as Partial Spoof (PS), can be more deceptive and challenging to detect.
Recent efforts to defend against such attacks have either extended utterance-level CMs to treat any manipulation as a global spoof~\cite{yi2022add, zhang2022partialspoof, wu2022partially}, or proposed frame-level detection methods (spoof localization) that identify “when” spoofing occurs within an utterance~\cite{yi21_interspeech, zhang2022localizing, liu24m_interspeech, zhong24_interspeech,TDL,GNCL}.

While spoof localization has enabled frame-level detection of spoofed regions within an utterance, it remains limited in that it does not provide information about how those frames were generated or which spoofing methods were used.
This lack of method-level insight hinders deeper forensic analysis, such as tracing the source of manipulation or profiling the type of attack.
Recognizing this incompleteness, Zhang et al.~\cite{Spoof_diarization} recently introduced a new task termed \textit{Spoof Diarization}, which builds upon spoof localization by not only identifying when spoofing occurs, but also clustering spoofed frames according to their spoofing methods—effectively answering the question of what spoofed and when.
This task opens new possibilities for forensic scenarios, particularly in source attribution and detection of unseen spoofing types.
As a first attempt to conduct spoof diarization, the authors proposed a baseline model composed of two separate branches: a localization branch that performs frame-level spoof detection, and a diarization branch that clusters spoofed frames based on their acoustic features to assign spoofing-type labels.
While this model laid the foundation for the task, its simple structure struggles to handle abrupt transitions between bona fide and spoofed frames and multiple attack types within a single utterance, limiting its adaptability to diverse spoofing scenarios.

To enhance spoof diarization beyond these initial limitations, we turn to speaker diarization, a closely related task that labels and clusters speaker-homogeneous segments to identify ``who spoke when."
Speaker diarization shares structural similarities with spoof diarization, as both tasks involve assigning latent class identities (speaker identities or spoof types) to audio frames under unknown class conditions and varying frame lengths.
Recent studies in speaker diarization have explored attractor-based modeling, a technique that uses reference vectors to represent class-specific centers in the embedding space~\cite{EEND-EDA, EEND-TA, AED-EEND, AED-EEND-EE, EEND-NAA, EEND-GLA}.
These reference vectors interact with frame-level embeddings to measure similarity, thereby guiding the model to capture class-discriminative information.
Through this mechanism, the model achieves strong performance in frame-level class assignment, even without access to explicit identity labels.

Inspired by these successes, we propose to integrate attractor-based modeling into spoof diarization by introducing learnable attractor tokens specifically tailored to represent bona fide and spoofed speech.
These tokens interact with frame-level embeddings via cross-attention, enabling simultaneous localization of spoofed frames and discrimination of spoofing methods.
By serving as reference points in the embedding space, the attractor tokens sharpen the boundary between bona fide and spoofed speech, while also capturing fine-grained variations among spoofing methods.
In addition to our attractor-based approach, we also construct a new baseline model by combining the localization and clustering branches into a single framework, aiming to reduce architectural complexity.
Our contributions are summarized as follows:
\begin{itemize}
\item To the best of our knowledge, this is the first study to incorporate attractor-based modeling from speaker diarization into the spoof diarization task.
\item We extend the evaluation to include spoof localization and utterance-level spoof detection, providing a more comprehensive view of the model’s behavior.
\item Results on the PartialSpoof dataset demonstrate that our proposed method achieves substantial improvements over the baseline, including a relative gain of 18.75\% in spoof diarization performance.
\end{itemize}

\section{Related Work} 
\subsection{Spoof Diarization}
\begin{figure}[htbp]
    \centerline{\includegraphics[width=1\columnwidth]{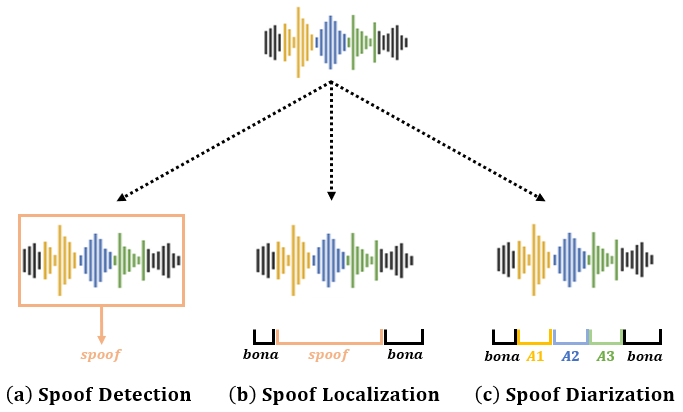}}
    \caption{Illustrations of (a) spoof detection, (b) spoof localization, and (c) spoof diarization. Black denotes bona fide speech (\textbf{bona}), ivory denotes all spoofed regions (\textbf{spoof}), additional colors indicate specific attack types (\textbf{A1}, \textbf{A2}, \textbf{A3}).}
    \label{fig:compare_spoof}
    \vspace{-0.4cm}
\end{figure}

\noindent
Spoof diarization, introduced in 2024, is a novel task that addresses the question of \textit{``what spoofed when"} in partially spoofed audio.
It extends the conventional goals of audio deepfake detection by integrating spoof detection, localization, and clustering of spoofing methods within a unified framework.
Specifically, spoof diarization generalizes prior tasks as follows (see Fig~\ref{fig:compare_spoof}):
\begin{itemize}
\item \textbf{Spoof detection}: Determines whether an utterance contains spoofed content via binary classification.
\item \textbf{Spoof localization}: Identifies which frames of the utterance are spoofed, without considering the spoofing method.
\item \textbf{Spoof diarization}: Assigns method-specific labels (e.g., A1, A2, A3) to spoofed frames while preserving bona fide labels, thereby producing a fine-grained, fully annotated timeline.
\end{itemize}

To realize spoof diarization in practice, a system must meet two core requirements.
First, it must segment the utterance into frame-level to accurately localize both bona fide and spoofed regions.
Second, it should be capable of distinguishing different spoofing methods, including those not seen during training, by assigning attack-specific labels to spoofed frames.
These dual objectives require a model to jointly perform temporal localization and spoof-type clustering under open-set conditions.

To satisfy these requirements, Zhang et al.~\cite{Spoof_diarization} proposed the Countermeasure-Condition Clustering (3C) model, which serves as the first benchmark for the spoof diarization task and as a baseline of this work.
The model adopts a dual-branch approach to handle both localization and clustering.

\subsection{The Baseline 3C Model}
\begin{figure*}[htbp]
    \centerline{\includegraphics[width=2\columnwidth]{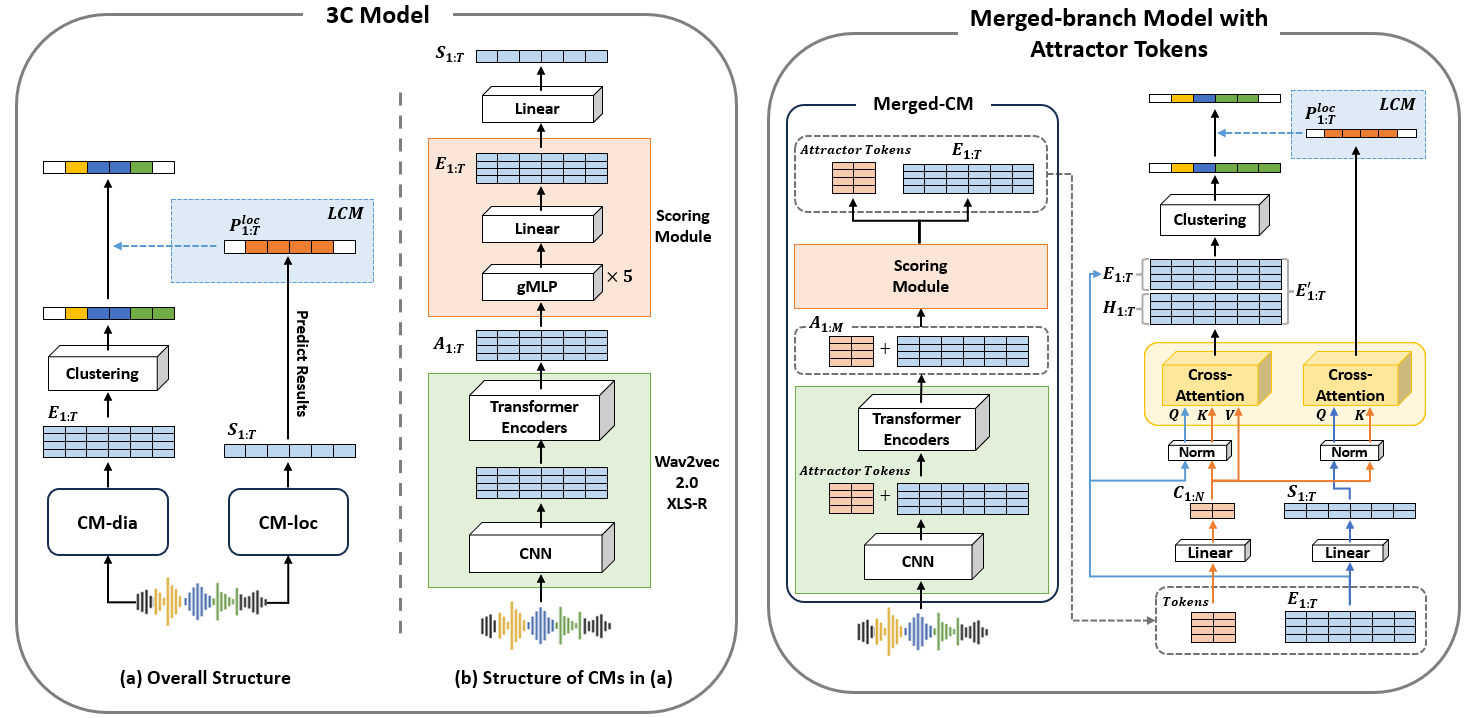}}
    \caption{The overall architecture of the baseline 3C model and our proposed Merged-branch model with attractor tokens. CM-dia and CM-loc represent the models for the diarization and localization branches, respectively. LCM denotes the label-based CM constraint module, and Norm indicates an $\ell_2$-normalization step.}
    \label{fig:3C model}
    \vspace{-0.3cm}
\end{figure*}

\noindent
As shown in the left part of Fig.\ref{fig:3C model}, the 3C model consists of two independent yet complementary branches: a diarization branch to cluster spoofing methods, and a localization branch to detect bona fide regions, enabling specialized and efficient handling of each subtask.

Both branches employ the Wav2Vec 2.0 XLS-R (W2V2)\cite{baevski2020wav2vec} as an acoustic front-end and utilize task-specific scoring modules as back-ends (see Fig.\ref{fig:3C model} b).
The diarization branch (CM-dia) is trained using a multi-class classification loss to distinguish among known spoofing methods, while the localization branch (CM-loc) uses binary classification to precisely separate bona fide from spoofed frames.
During inference, CM-dia extracts frame-level embeddings $E_{1:T}$ from the input audio, which are clustered to assign method-specific labels to spoofed frames.
In parallel, CM-loc produces frame-wise scores $S_{1:T}$ indicating the likelihood of each frame being bona fide.
The outputs of both branches are integrated using a label-based CM constraint (LCM) module, which preserves bona fide predictions from CM-loc when generating the final sequence of spoof-type and bona fide labels.

This dual-branch structure allows CM-dia to cluster spoofed frames based on acoustic patterns, including the discovery of novel spoofing methods, whereas CM-loc ensures accurate temporal localization between spoofed and bona fide speech.

\subsection{Attractor Mechanisms in Speaker Diarization}
\noindent
In speaker diarization, attractors are learnable reference vectors that serve as anchor points for associating frame-level features with speaker identities. 
They are particularly effective in open-set conditions, where the number and identity of speakers vary across utterances. 
Recent studies have implemented attractors using diverse neural architectures. 
Horiguchi et al.~\cite{EEND-EDA} introduced EEND-EDA, employing an LSTM encoder-decoder to dynamically generate speaker-specific attractors, thus removing fixed speaker count limitations. 
Samarakoon et al.\cite{EEND-TA} proposed EEND-TA, which utilizes an attention-based Transformer decoder instead of an LSTM, resulting in more discriminative speaker representations.
Meanwhile, Chen et al.~\cite{AED-EEND-EE} approached the problem differently with AED-EEND-EE, focusing on the interaction between frame-level features and speaker attractors. 
They applied cross-attention mechanisms, where attractors function as targeted queries against frame embeddings, yielding highly distinctive speaker characteristics.
Overall, these approaches highlight the versatility of attractor-based modeling in speaker diarization and demonstrate its effectiveness in improving segmentation and clustering performance under challenging conditions.

Spoof diarization shares a structural similarity with speaker diarization in that both tasks require frame-level assignment to latent classes—speaker identities in the former, and spoofing methods or bona fide labels in the latter—without prior knowledge of the number or types of classes involved. 
Given this parallel task structure, attractor-based modeling can be similarly applied to spoof diarization, where learned reference vectors could guide the clustering of spoofed frames and aid in distinguishing them from bona fide speech.

\section{Proposed Method}
\noindent
We propose to incorporate an attractor mechanism into the spoof diarization task. 
Our approach encourages the model to learn distinct representations for bona fide and spoofed speech, guiding the model to form distinguishable feature spaces for each spoofing type. 
In this section, we describe the detailed model architecture and training objectives.

\subsection{Model Architecture}
\noindent
We first introduce a novel method that integrates learnable attractor tokens, aiming to enhance discriminative representations for bona fide and spoofed speech. 
While the original 3C model employs two independent branches (diarization and localization), we primarily demonstrate our attractor-based modeling using a simplified merged-branch architecture for clarity and computational efficiency. 
Note that our attractor token method is model-agnostic and can be seamlessly integrated into various architectures, including the original 3C model.

The right part of Fig.\ref{fig:3C model} illustrates how the attractor tokens are integrated into the model to interact with frame-level features.
The proposed merged-branch architecture begins by extracting latent features from the CNN encoder of Wav2Vec 2.0 XLS-R (W2V2), to which two learnable attractor tokens are subsequently appended.  
These tokens, along with the frame-level features, are passed through a Transformer encoder to capture complex temporal dependencies.  
To highlight salient information, we then apply a weighted summation over the outputs of all Transformer layers, separately for the frame-level features and the attractor tokens.  
The resulting aggregated features are concatenated along the time axis, forming a unified sequence \( A_{1:M} \in \mathbb{R}^{M \times d} \), where \( M = T + N \); here, \( T \) is the number of frame-level embeddings, \( N = 2 \) denotes the number of attractor tokens, and \( d \) is the feature dimension.

Once concatenated, the sequence \( A_{1:M} \) is passed through a gMLP layer~\cite{liu2021pay}, denoted as \( g(\cdot) \), which captures both local and global dependencies within the sequence:
\begin{equation}
E_{1:M} = g(A_{1:M}),
\end{equation}
where \( E_{1:M} \in \mathbb{R}^{M \times d} \) represents the transformed embeddings.  
This output is then partitioned into frame-level representations \( E_{1:T} \in \mathbb{R}^{T \times d} \) and token embeddings \( E_{T:M} \in \mathbb{R}^{N \times d} \), corresponding to their respective input types.
Each is passed through a separate linear projection to prepare the features for their respective downstream roles:
\begin{align}
S_{1:T} &= \phi(E_{1:T} W_s + b_s), \\
C_{1:N} &= \phi(E_{T:M} W_c + b_c),
\end{align}
where \( W_s, W_c \in \mathbb{R}^{d \times d} \) and \( b_s, b_c \in \mathbb{R}^{d} \) are trainable parameters, and \( \phi(\cdot) \) denotes the GELU activation function~\cite{hendrycks2016gaussian}.

To guide the model in differentiating between bona fide and spoofed, we apply cross-attention between the frame-level representations and the two attractor tokens. 
First, we compute the attention weights between the normalized frame-level projection $\tilde{S}_{1:T}$ and token embeddings $\tilde{C}_{1:N}$ as:
\begin{equation}
P_{1:T}^{\text{loc}} = \text{softmax}(\frac{(\tilde{S}_{1:T} W_{Q_s})(\tilde{C}_{1:N} W_{K_s})^\top}{\sqrt{d}}),
\end{equation}
where $W_{Q_s}, W_{K_s} \in \mathbb{R}^{d \times d}$ are learnable projection matrices, and the softmax is applied along the key (token) dimension to produce an attention map of shape $T \times N$. 
This attention map $P_{1:T}^{\text{loc}}$ reflects the degree of alignment between each frame and the attractor tokens, enabling the model to explicitly highlight frames that exhibit characteristics indicative of bona fide or spoofed speech.

Next, a separate attention mechanism is employed to compute frame-level embeddings enhanced via the tokens.
Specifically, we compute:
\begin{equation}
H_{1:T} = \text{softmax}(\frac{(\tilde{E}_{1:T} W_{Q_e})(\tilde{C}_{1:N} W_{K_e})^\top}{\sqrt{d}})(C_{1:N} W_{V_e}),
\end{equation}
where $W_{Q_e}, W_{K_e}, W_{V_e} \in \mathbb{R}^{d \times d}$ are projection matrices for the second attention block. 
In this formulation, $C_{1:N}$ serves not only as the key but also as the value, allowing spoof type-specific information to be injected into each frame representation. 
The resulting features $H_{1:T}$ encode features enriched with spoofing-related context from the attractor tokens, serving as complementary information to the original features.

Finally, we obtain the final frame-level embeddings ($E'_{1:T}$) by concatenating the original frame-level embeddings with the attractor-conditioned features along the feature dimension:
\begin{equation}
    E'_{1:T} = \text{Concat}(E_{1:T}, H_{1:T}).
\end{equation}
These embeddings contain not only frame-specific information but also relational cues to the spoofing patterns present in the entire utterance. 
Through this enriched representation, the model can perform clustering and classification based on more discriminative and informative features.

\subsection{Training Objectives}
\noindent
To optimize the spoof diarization model, we employ the probability-to-similarity gradient (P2SGrad) loss \cite{zhang2019p2sgrad}, a technique previously shown to be effective in related tasks. 
In contrast to previous work, we extend its application beyond frame-level predictions by incorporating the learnable tokens, and modify the computation of $P_{1:T}^{\text{loc}}$ to align with our attractor-based architecture.

The P2SGrad defines its objective directly in the cosine similarity space by comparing $\ell_2$-normalized embeddings and class prototypes. 
Specifically, it computes the mean squared error between these similarity scores and one-hot target labels.
Given a normalized learnable class prototype $\tilde{o}_k$ and a normalized embedding $\tilde{x}_i$, the cosine similarity is defined as:
\begin{equation}
P_{i,k} = \tilde{x}_i^\top \tilde{o}_k,
\end{equation}
where $i$ indexes each frame in the current mini-batch and $k \in \{1, \dots, L\}$ indexes one of the $L$ target classes.
The P2SGrad loss is then computed across a mini-batch of $B$ utterances, each with $T$ frames, yielding a total of $B \times T$ frame-level embeddings. 
For each embedding, the similarity with all class prototypes is evaluated, and the loss is formulated as:
\begin{equation}
\mathcal{L}_{\text{P2SGrad}}(P,y) = \frac{1}{|B| \times |T|} \sum_{i=1}^{|B| \times |T|} \sum_{k=1}^{L} \left( P_{i,k} - \indicator{y_i = k} \right)^2,
\end{equation}
where $\indicator{y_i = k}$ is 1 if the label of sample $i$ equals $k$, and 0 otherwise.
This formulation aligns cosine similarity scores with discrete class labels in a regression-based manner.
In doing so, it encourages each embedding to move closer to its target class prototype while staying distant from non-target classes.

To support both frame-level and token-level supervision, we compute prediction scores from frame-level embeddings $E'_{1:T}$ and token-level embeddings $C_{1:N}$, by comparing them with learnable class-specific prototypes $O_{\text{dia}}$ and $O_{\text{token}}$ in $\mathbb{R}^{d \times L}$:
\begin{align}
P_{1:T}^{\text{dia}} &= \tilde{E}'_{1:T} \tilde{O}_{\text{dia}}, \\
P^{\text{token}} &= \text{Concat}(  \tilde{C}_{1} \tilde{O}_{\text{token}_1}, \tilde{C}_{2} \tilde{O}_{\text{token}_2}),
\end{align}
where $O_{\text{token}_1} \in \mathbb{R}^{d \times 1}$ and $O_{\text{token}_2} \in \mathbb{R}^{d \times (L-1)}$ represent the bona fide and spoofed prototypes, respectively.
The resulting similarity matrices are $P_{1:T}^{\text{dia}} \in \mathbb{R}^{T \times L}$ and $P^{\text{token}} \in \mathbb{R}^{1 \times L}$.

Based on these predictions, we formulate three loss terms using P2SGrad:
\begin{align}
\mathcal{L}_{\text{loc}} &= \mathcal{L}_{\text{P2SGrad}}(P_{1:T}^{\text{loc}}, y^{\text{loc}}), \\
\mathcal{L}_{\text{dia}} &= \mathcal{L}_{\text{P2SGrad}}(P_{1:T}^{\text{dia}}, y^{\text{dia}}), \\
\mathcal{L}_{\text{token}} &= \mathcal{L}_{\text{P2SGrad}}(P^{\text{token}}, y^{\text{token}}).
\end{align}
Here, \( y^{\text{loc}} \) is a binary frame-level label that indicates whether each frame is bona fide or spoofed, while \( y^{\text{dia}} \) corresponds to a frame-level one-hot vector denoting the specific spoofing type (or bona fide) present at each frame.  
In contrast, \( y^{\text{token}} \) is a multi-label utterance-level vector that encodes the presence of all spoofing types (including bona fide) within the utterance, providing global supervision for the token representations.

Finally, the total loss is defined as:
\begin{equation}
\mathcal{L}_{\text{total}} = \mathcal{L}_{\text{loc}} + \mathcal{L}_{\text{dia}} + \mathcal{L}_{\text{token}}.
\end{equation}
By jointly optimizing the three objectives, the model learns to distinguish bona fide frames from spoofed ones and to classify the spoofed frames according to their attack types.  
The tokens, in turn, capture utterance-level spoofing characteristics, allowing them to function as attractors.

\section{Experimental Setup}
\subsection{Datasets and Implementation Details}
\noindent
We conducted our experiments primarily on the PartialSpoof dataset~\cite{wu2022partially}, which provides a balanced distribution of spoofing methods across its training, development, and evaluation partitions. 
The proportion of bona fide speech in each set is 55.3\%, 56.0\%, and 60.7\%, respectively. 
To further assess utterance-level spoof detection performance, we also evaluated our models using the ASVspoof 2019 LA~\cite{yamagishi2019asvspoof}, ASVspoof2021 LA and DF ~\cite{delgado2021asvspoof} datasets. 
During evaluation, non-speech frames were excluded from scoring to focus solely on speech-active regions. 
In contrast, for training, we retained silences only in the “concatenated-parts” subset, where spoofed and bona fide frames are joined. 
All countermeasure modules were operated with a time resolution of 20 ms. 
A binary decision threshold of 0.5 was used for classification, selected based on empirical tuning.
For clustering, we employed agglomerative hierarchical clustering using cosine distance. 
Following previous evaluations, we utilized the ground-truth number of clusters (oracle setting) to isolate clustering performance from cluster estimation errors.
The best-performing models on the development set were used for evaluation.

\subsection{Evaluation Metrics}
\noindent
To evaluate both frame-level spoof detection and spoof-type clustering, we adopt two Jaccard-based metrics proposed by Zhang et al.~\cite{Spoof_diarization}: $\textnormal{JI}_\textnormal{bona}$ and $\textnormal{JER}_\textnormal{spoof}$. 
$\textnormal{JI}_\textnormal{bona}$ measures the detection quality of bona fide frames by accounting for false alarms and missed detections, while $\textnormal{JER}_\textnormal{spoof}$ quantifies clustering errors across spoof types by averaging Jaccard error. 

The global scores for both metrics are defined as:
\begin{equation}
\text{JI}_{\text{bona}} = \frac{1}{|D|} \sum_{j \in D} \frac{\text{FA}_{\text{bona}, j} + \text{MD}_{\text{bona}, j}}{\text{TOTAL}_{\text{bona}, j}}
\end{equation}
\vspace{-0.1cm}
\begin{equation}
\text{JER}_{\text{spoof}} = \frac{1}{\sum_{j \in D} |A_j|} \sum_{j \in D} \sum_{A_i \in A_j} \frac{\text{FA}_{A_i, j} + \text{MD}_{A_i, j}}{\text{TOTAL}_{A_i, j}}
\end{equation}
where $D$ denotes the set of all evaluation utterances, and $A_j$ denotes the set of spoofing methods appearing in utterance $j$.
The symbol $| \cdot |$ indicates set cardinality.
For each label $k \in \{\text{bona}, A_i\}$, $\text{FA}_{k, j}$ and $\text{MD}_{k, j}$ indicate the durations of false alarms and missed detections in utterance $j$, while $\text{TOTAL}_{k, j}$ corresponds to the total duration covered by the union of the reference and prediction frames for label $k$.

\section{Results}
\noindent
Table~\ref{tab:main} compares the baseline model with our proposed attractor-based method on the PartialSpoof development and evaluation sets.
To verify the general effectiveness of the proposed attractor tokens, we apply them not only to the original dual-branch 3C model but also to a simplified Merged-branch model, which merges the diarization and localization branches.
We first evaluate our own implementation of the 3C model (denoted as 3C model*), and then assess the effect of applying the attractor token to this baseline. 
The results show that the attractor token brings clear gains, with relative improvements of 18.6\% in $\textnormal{JI}_\textnormal{bona}$ and 13.67\% in $\textnormal{JER}_\textnormal{spoof}$ on the evaluation set.
Next, we examine the Merged-branch model that merges the CM-Dia and CM-Loc branches to reduce training cost and halve the model size. 
While this simplified configuration shows a moderate drop in performance compared to 3C model*, it still provides results close to the baseline. 
With the attractor token applied, the Merged-branch model achieves further relative improvements of 20.3\% and 17.2\%, respectively.
These results confirm that the attractor tokens consistently enhance spoof diarization performance across different model architectures by improving the separation between bona fide and spoofed regions and promoting discriminative clustering, even under unseen attack conditions.

\begin{table}[t]
\centering
\caption{Spoof diarization results on the PartialSpoof database, showing the effect of attractor tokens across different model configurations. Bold values indicate the best results; underlined ones show improvements achieved by our method.}
\label{tab:main}
\renewcommand{\arraystretch}{1.2}
\resizebox{\linewidth}{!}{%
\begin{tabular}{l|cc|cc}
\Xhline{3\arrayrulewidth}
\multirow{2}{*}{Model} & \multicolumn{2}{c|}{Development set} & \multicolumn{2}{c}{Evaluation set} \\ \cline{2-5}
& \textnormal{JI\textsubscript{bona}} (\%) & \textnormal{JER\textsubscript{spoof}} (\%) & \textnormal{JI\textsubscript{bona}} (\%) & \textnormal{JER\textsubscript{spoof}} (\%) \\
\hline
3C model \cite{Spoof_diarization} & 4.49 & 5.27 & \textbf{15.15} & 34.13 \\ \hline
3C model* & \textbf{3.15} & \textbf{3.76} & 19.13 & 28.24 \\
\quad w/ attractor tokens & 3.35 & 4.36 & \uline{15.58} & \textbf{24.38} \\ \hline
Merged-branch model & 3.37 & 4.11 & 22.64 & 32.59 \\
\quad w/ attractor tokens & \uline{3.23} & \uline{3.85} & \uline{18.05} & \uline{26.99} \\
\Xhline{3\arrayrulewidth}
\end{tabular}%
}
\end{table}

\begin{figure}[t!]
    \centerline{\includegraphics[width=1\columnwidth]{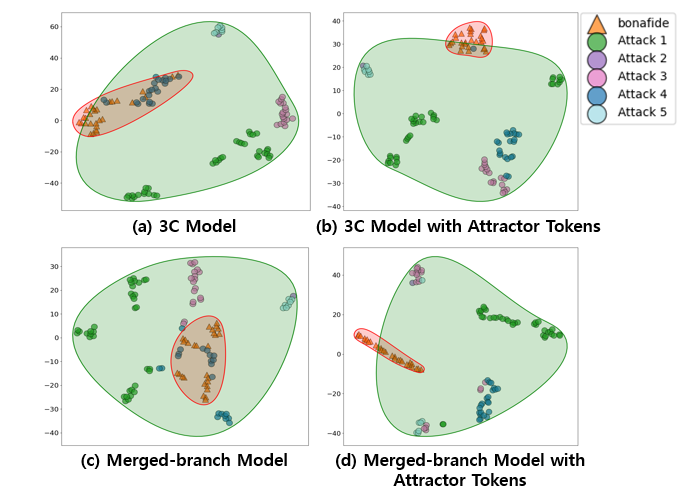}}
    \caption{Visualization of frame-level embeddings via t-SNE on the PartialSpoof evaluation set. The embeddings are extracted from the 3C and Merged-branch models, each with and without attractor tokens. Green and red regions represent clusters of spoofed and bona fide embeddings, respectively.}
    \label{fig:visualization}
    \vspace{-0.4cm}
\end{figure}

Fig.\ref{fig:visualization} illustrates the distribution of frame-level embeddings when attractor tokens are applied to the 3C model (a) and the Merged-branch model (c).
In both (a) and (c), numerous spoof-type embeddings are observed within the bona fide region, indicating insufficient separation.
In contrast, the attractor token-applied versions, shown in (b) and (d), exhibit more clearly defined bona fide regions and substantially stronger clustering effects.
These results suggest that the proposed attractor tokens not only improve the separation between bona fide and spoofed regions but also enhance the discriminability among different spoofing types.

\begin{table}[t]
\centering
\caption{Comparison of label schemes for diarization (‘Dia’) and localization (‘Loc’) in 3C (branch-based) and Merged-branch (objective-based) models. ‘Mul’ includes bona fide and multiple spoof types; ‘Spf’ uses spoof-only labels; ‘Bin’ applies binary supervision. The ‘Mul+Bin’ settings match the configuration in Table~\ref{tab:main}.}
\label{tab:label}
\renewcommand{\arraystretch}{1.2}
\resizebox{\linewidth}{!}{%
\begin{tabular}{l|cc|cc|cc}
\Xhline{3\arrayrulewidth}
\multirow{2}{*}{Model} & \multicolumn{2}{c|}{Label} & \multicolumn{2}{c|}{Dev set} & \multicolumn{2}{c}{Eval set} \\ \cline{2-7}
& Dia. & Loc. & \textnormal{JI\textsubscript{bona}} & \textnormal{JER\textsubscript{spoof}} & \textnormal{JI\textsubscript{bona}} & \textnormal{JER\textsubscript{spoof}} \\
\hline
\multirow{3}{*}{3C model \cite{Spoof_diarization}} & Mul & Bin & 4.49 & 5.27 & \textbf{15.15} & 34.13 \\
& Spf & Bin & 4.52 & 5.71 & 15.18 & 36.03 \\
& Mul & / & 4.49 & 5.21 & 19.66 & 28.05 \\ \hline
\multirow{3}{*}{Merged-branch model} & Mul & Bin & 3.37 & 4.11 & 22.64 & 32.59 \\
& Spf & Bin & 4.12 & 10.81 & 19.95 & 31.73 \\
& Mul & / & 3.73 & 4.51 & 22.83 & 32.17 \\ \hline
\multirow{3}{*}{\quad w/ attractor tokens} & Mul & Bin & \textbf{3.23} & \textbf{3.85} & 18.05 & \textbf{26.99} \\
& Spf & Bin & 3.66 & 7.34 & 17.34 & 27.05 \\
& Mul & / & 3.56 & 4.26 & 22.01 & 31.80 \\
\Xhline{3\arrayrulewidth}
\end{tabular}%
}
\vspace{-0.4cm}
\end{table}

To assess the robustness of the attractor token under different label configurations, we varied the supervision used for both the diarization and localization objectives within the Merged-branch model.
For the diarization objective, we tested two schemes: ‘Mul’, which distinguishes among bona fide, concatenation parts (i.e., frames where speech from different classes are seamlessly joined), and multiple spoof types; and ‘Spf’, which includes only spoof-type labels and excludes bona fide.
For the localization objective, we either applied binary labels distinguishing bona fide and spoofed speech (denoted as ‘Bin’) or removed the corresponding loss term $\mathcal{L}_{\text{loc}}$.
As shown in Table~\ref{tab:label}, the Merged-branch model maintained stable performance across label settings.
With the attractor token, the ‘Spf+Bin’ configuration achieved relative improvements of 13.1\% in $\textnormal{JI}_\textnormal{bona}$ and 14.7\% in $\textnormal{JER}_\textnormal{spoof}$ compared to the baseline. 
In contrast, removing the binary classification loss weakened the impact of the attractor token, indicating that its effectiveness relies on complementary binary supervision.
This suggests that the token benefits not only from the token-level representation learning, but also from being guided by explicit frame-wise bona fide/spoof discrimination signals.
Overall, the attractor token consistently improved spoof diarization performance across varying label configurations, indicating its potential to reduce sensitivity to label design choices in training.

To analyze how the attractor token influences different parts of the spoof diarization system, we conducted an ablation study by selectively applying the token to the diarization and localization objectives.
As shown in Table~\ref{tab:ablation}, applying the token to either objective individually led to performance improvements over the baseline, suggesting that the token plays a meaningful role in both cases.
When the token was applied only to the localization objective (Loc), the model exhibited consistent gains, which suggests that the token learned spoof- and bona fide-related information distributed across an utterance and leveraged it to support binary frame-level classification. 
This led to more effective separation of the two classes compared to the baseline.
In contrast, applying the token only to the diarization objective (Dia) likely benefited from the token capturing class-specific characteristics and interacting with frame-level embeddings through cross-attention, thereby facilitating more discriminative representations for clustering.
The best overall results were obtained when the token was applied to both objectives, indicating that the attractor token effectively adapts to their distinct roles and leverages complementary supervision signals to enhance spoof diarization performance.

Table~\ref{tab:utt} presents spoof detection results at both the utterance and frame levels.
While the overall performance is not competitive with existing spoof detection or localization systems, the attractor token still leads to clear improvements over the baseline.
This reflects the potential of the attractor token to contribute beyond spoof diarization.
However, the results also reveal a clear need for spoof diarization to develop in ways that can better complement and integrate with the objectives of spoof detection and localization research.

\begin{table}[t]
\centering
\caption{Ablation results on the PartialSpoof database with different token guidance configurations. A check mark (\checkmark) indicates the objective where the attractor token was applied.}
\label{tab:ablation}
\renewcommand{\arraystretch}{1.2}
\resizebox{\linewidth}{!}{%
\begin{tabular}{c|cc|cc|cc}
\Xhline{3\arrayrulewidth}
\multirow{2}{*}{Model} & \multicolumn{2}{c|}{Objectives} & \multicolumn{2}{c|}{Dev set} & \multicolumn{2}{c}{Eval set} \\ \cline{2-7}
& Dia. & Loc. & \textnormal{JI\textsubscript{bona}} & \textnormal{JER\textsubscript{spoof}} & \textnormal{JI\textsubscript{bona}} & \textnormal{JER\textsubscript{spoof}} \\
\hline
Merged-branch model & -- & -- & 3.37 & 4.11 & 22.64 & 32.59 \\ \hline
\quad w/ attractor tokens & -- & \checkmark & \textbf{3.19} & \textbf{3.83} & 20.46 & 29.79 \\
\quad w/ attractor tokens & \checkmark & -- & 3.27 & 4.06 & 21.22 & 29.92 \\
\quad w/ attractor tokens & \checkmark & \checkmark & 3.23 & 3.85 & \textbf{18.05} & \textbf{26.99} \\
\Xhline{3\arrayrulewidth}
\end{tabular}%
}
\end{table}

\begin{table}[t]
\centering
\caption{
Utterance-level and frame-level spoof detection results on the PartialSpoof (PS), ASVspoof 2019 LA (19LA), 2021 LA (21LA) and DF eval sets (21DF).
}
\label{tab:utt}
\renewcommand{\arraystretch}{1.2}
\resizebox{0.9\linewidth}{!}{%
\begin{tabular}{l|c|c|c|c|c}
\Xhline{3\arrayrulewidth}
\multirow{2}{*}{Model} & Frame. EER (\%) & \multicolumn{4}{c}{Utt. EER (\%)} \\ \cline{2-6}
& PS & PS & 19LA & 21LA & 21DF\\ \hline
3C model* & 41.84 & 43.60 & 39.01 & 41.97 & 36.62\\
\quad w/ attractor tokens & 21.26 & \textbf{14.02} & 31.72 & 33.11 & 31.93 \\ \hline
Merged-branch model & 21.91 & 26.14 & 38.64 & 36.49 & 31.84 \\
\quad w/ attractor tokens & \textbf{19.80} & 18.23 & \textbf{27.40} & \textbf{19.65} & \textbf{27.38} \\
\Xhline{3\arrayrulewidth}
\end{tabular}%
}
\vspace{-0.4cm}
\end{table}

\section{Conclusion}
\noindent
This paper presents an attractor token-based method for the spoof diarization task, which aims to determine ``what spoofed when'' in partially manipulated utterances. 
The attractor tokens enhance the model’s ability to distinguish bona fide and spoofed speech by interacting with frame-level embeddings through cross-attention to extract discriminative patterns.
We applied the proposed approach to both the original dual-branch architecture (3C model) and a simplified Merged-branch model. 
Experiments on the PartialSpoof dataset show consistent improvements in both settings, with up to 20.3\% and 17.2\% relative gains in JI$_{\text{bona}}$ and JER$_{\text{spoof}}$, respectively, demonstrating the effectiveness of attractor-based modeling for spoof diarization.
This approach can lay the groundwork for future spoof detection systems that require both fine-grained localization and semantic spoof analysis.
However, performance degradation under unseen spoofing attacks and limited robustness at the utterance level remain unresolved issues.
Future work will focus on addressing these challenges.

\newpage


\bibliographystyle{IEEEtran}
\bibliography{refs}

\begin{thebibliography}{10}
\providecommand{\url}[1]{#1}
\csname url@samestyle\endcsname
\providecommand{\newblock}{\relax}
\providecommand{\bibinfo}[2]{#2}
\providecommand{\BIBentrySTDinterwordspacing}{\spaceskip=0pt\relax}
\providecommand{\BIBentryALTinterwordstretchfactor}{4}
\providecommand{\BIBentryALTinterwordspacing}{\spaceskip=\fontdimen2\font plus
\BIBentryALTinterwordstretchfactor\fontdimen3\font minus \fontdimen4\font\relax}
\providecommand{\BIBforeignlanguage}[2]{{%
\expandafter\ifx\csname l@#1\endcsname\relax
\typeout{** WARNING: IEEEtran.bst: No hyphenation pattern has been}%
\typeout{** loaded for the language `#1'. Using the pattern for}%
\typeout{** the default language instead.}%
\else
\language=\csname l@#1\endcsname
\fi
#2}}
\providecommand{\BIBdecl}{\relax}
\BIBdecl

\bibitem{triantafyllopoulos2023overview}
A.~Triantafyllopoulos, B.~W. Schuller, G.~{\.I}ymen, M.~Sezgin, X.~He, Z.~Yang, P.~Tzirakis, S.~Liu, S.~Mertes, E.~Andr{\'e} \emph{et~al.}, ``An overview of affective speech synthesis and conversion in the deep learning era,'' \emph{Proceedings of the IEEE}, vol. 111, no.~10, pp. 1355--1381, 2023.

\bibitem{li2023styletts}
Y.~A. Li, C.~Han, V.~Raghavan, G.~Mischler, and N.~Mesgarani, ``Styletts 2: Towards human-level text-to-speech through style diffusion and adversarial training with large speech language models,'' \emph{Advances in Neural Information Processing Systems}, vol.~36, pp. 19\,594--19\,621, 2023.

\bibitem{tan2024naturalspeech}
X.~Tan, J.~Chen, H.~Liu, J.~Cong, C.~Zhang, Y.~Liu, X.~Wang, Y.~Leng, Y.~Yi, L.~He \emph{et~al.}, ``Naturalspeech: End-to-end text-to-speech synthesis with human-level quality,'' \emph{IEEE Transactions on Pattern Analysis and Machine Intelligence}, vol.~46, no.~6, pp. 4234--4245, 2024.

\bibitem{wu2015asvspoof}
Z.~Wu, T.~Kinnunen, N.~Evans, J.~Yamagishi, C.~Hanil{\c{c}}i, M.~Sahidullah, and A.~Sizov, ``Asvspoof 2015: the first automatic speaker verification spoofing and countermeasures challenge,'' in \emph{INTERSPEECH 2015, Automatic Speaker Verification Spoofing and Countermeasures Challenge, colocated with INTERSPEECH 2015}.\hskip 1em plus 0.5em minus 0.4em\relax ISCA, 2015, pp. 2037--2041.

\bibitem{kinnunen2017asvspoof}
T.~Kinnunen, M.~Sahidullah, H.~Delgado, M.~Todisco, N.~Evans, J.~Yamagishi, and K.~A. Lee, ``The asvspoof 2017 challenge: Assessing the limits of replay spoofing attack detection,'' 2017.

\bibitem{yamagishi2019asvspoof}
J.~Yamagishi, M.~Todisco, M.~Sahidullah, H.~Delgado, X.~Wang, N.~Evans, T.~Kinnunen, K.~A. Lee, V.~Vestman, and A.~Nautsch, ``Asvspoof 2019: The 3rd automatic speaker verification spoofing and countermeasures challenge database,'' 2019.

\bibitem{delgado2021asvspoof}
H.~Delgado, N.~Evans, T.~Kinnunen, K.~A. Lee, X.~Liu, A.~Nautsch, J.~Patino, M.~Sahidullah, M.~Todisco, X.~Wang \emph{et~al.}, ``Asvspoof 2021: Automatic speaker verification spoofing and countermeasures challenge evaluation plan,'' \emph{arXiv preprint arXiv:2109.00535}, 2021.

\bibitem{wang2025asvspoof}
X.~Wang, H.~Delgado, H.~Tak, J.-w. Jung, H.-j. Shim, M.~Todisco, I.~Kukanov, X.~Liu, M.~Sahidullah, T.~Kinnunen \emph{et~al.}, ``Asvspoof 5: Design, collection and validation of resources for spoofing, deepfake, and adversarial attack detection using crowdsourced speech,'' \emph{arXiv preprint arXiv:2502.08857}, 2025.

\bibitem{jung2022aasist}
J.-w. Jung, H.-S. Heo, H.~Tak, H.-j. Shim, J.~S. Chung, B.-J. Lee, H.-J. Yu, and N.~Evans, ``Aasist: Audio anti-spoofing using integrated spectro-temporal graph attention networks,'' in \emph{ICASSP 2022-2022 IEEE international conference on acoustics, speech and signal processing (ICASSP)}.\hskip 1em plus 0.5em minus 0.4em\relax IEEE, 2022, pp. 6367--6371.

\bibitem{tak2021end}
H.~Tak, J.~Patino, M.~Todisco, A.~Nautsch, N.~Evans, and A.~Larcher, ``End-to-end anti-spoofing with rawnet2,'' in \emph{ICASSP 2021-2021 IEEE International Conference on Acoustics, Speech and Signal Processing (ICASSP)}.\hskip 1em plus 0.5em minus 0.4em\relax IEEE, 2021, pp. 6369--6373.

\bibitem{shin2024hm}
H.-s. Shin, J.~Heo, J.-h. Kim, C.-y. Lim, W.~Kim, and H.-J. Yu, ``Hm-conformer: A conformer-based audio deepfake detection system with hierarchical pooling and multi-level classification token aggregation methods,'' in \emph{ICASSP 2024-2024 IEEE International Conference on Acoustics, Speech and Signal Processing (ICASSP)}.\hskip 1em plus 0.5em minus 0.4em\relax IEEE, 2024, pp. 10\,581--10\,585.

\bibitem{wu24b_interspeech}
H.~Wu, W.~Guo, Z.~Zhang, W.~Zhao, S.~Peng, and J.~Zhang, ``Spoofing speech detection by modeling local spectro-temporal and long-term dependency,'' in \emph{Interspeech 2024}, 2024, pp. 507--511.

\bibitem{STCA}
Y.~Hao, M.~Xu, Y.~Chen, Y.~Liu, L.~He, L.~Fang, and L.~Liu, ``Integrating spectro-temporal cross aggregation and multi-scale dynamic learning for audio deepfake detection,'' in \emph{ICASSP 2025 - 2025 IEEE International Conference on Acoustics, Speech and Signal Processing (ICASSP)}, 2025, pp. 1--5.

\bibitem{CMFF}
Z.~Jin, L.~Lang, and B.~Leng, ``Wave-spectrogram cross-modal aggregation for audio deepfake detection,'' in \emph{ICASSP 2025 - 2025 IEEE International Conference on Acoustics, Speech and Signal Processing (ICASSP)}, 2025, pp. 1--5.

\bibitem{yi2022add}
J.~Yi, R.~Fu, J.~Tao, S.~Nie, H.~Ma, C.~Wang, T.~Wang, Z.~Tian, Y.~Bai, C.~Fan \emph{et~al.}, ``Add 2022: the first audio deep synthesis detection challenge,'' in \emph{ICASSP 2022-2022 IEEE International Conference on Acoustics, Speech and Signal Processing (ICASSP)}.\hskip 1em plus 0.5em minus 0.4em\relax IEEE, 2022, pp. 9216--9220.

\bibitem{zhang2022partialspoof}
L.~Zhang, X.~Wang, E.~Cooper, N.~Evans, and J.~Yamagishi, ``The partialspoof database and countermeasures for the detection of short fake speech segments embedded in an utterance,'' \emph{IEEE/ACM Transactions on Audio, Speech, and Language Processing}, vol.~31, pp. 813--825, 2022.

\bibitem{wu2022partially}
H.~Wu, H.-C. Kuo, N.~Zheng, K.-H. Hung, H.-Y. Lee, Y.~Tsao, H.-M. Wang, and H.~Meng, ``Partially fake audio detection by self-attention-based fake span discovery,'' in \emph{ICASSP 2022-2022 IEEE International Conference on Acoustics, Speech and Signal Processing (ICASSP)}.\hskip 1em plus 0.5em minus 0.4em\relax IEEE, 2022, pp. 9236--9240.

\bibitem{yi21_interspeech}
J.~Yi, Y.~Bai, J.~Tao, H.~Ma, Z.~Tian, C.~Wang, T.~Wang, and R.~Fu, ``Half-truth: A partially fake audio detection dataset,'' in \emph{Interspeech 2021}, 2021, pp. 1654--1658.

\bibitem{zhang2022localizing}
B.~Zhang and T.~Sim, ``Localizing fake segments in speech,'' in \emph{2022 26th International Conference on Pattern Recognition (ICPR)}.\hskip 1em plus 0.5em minus 0.4em\relax IEEE, 2022, pp. 3224--3230.

\bibitem{liu24m_interspeech}
T.~Liu, L.~Zhang, R.~K. Das, Y.~Ma, R.~Tao, and H.~Li, ``How do neural spoofing countermeasures detect partially spoofed audio?'' in \emph{Interspeech 2024}, 2024, pp. 1105--1109.

\bibitem{zhong24_interspeech}
J.~Zhong, B.~Li, and J.~Yi, ``Enhancing partially spoofed audio localization with boundary-aware attention mechanism,'' in \emph{Interspeech 2024}, 2024, pp. 4838--4842.

\bibitem{TDL}
Y.~Xie, H.~Cheng, Y.~Wang, and L.~Ye, ``An efficient temporary deepfake location approach based embeddings for partially spoofed audio detection,'' in \emph{ICASSP 2024 - 2024 IEEE International Conference on Acoustics, Speech and Signal Processing (ICASSP)}, 2024, pp. 966--970.

\bibitem{GNCL}
Z.~Ge, X.~Xu, H.~Guo, Z.~Yang, and B.~Schuller, ``Gncl: A graph neural network with consistency loss for segment-level spoofed speech detection,'' in \emph{ICASSP 2025 - 2025 IEEE International Conference on Acoustics, Speech and Signal Processing (ICASSP)}, 2025, pp. 1--5.

\bibitem{Spoof_diarization}
L.~Zhang, X.~Wang, E.~Cooper, M.~Diez, F.~Landini, N.~Evans, and J.~Yamagishi, ``Spoof diarization: "what spoofed when" in partially spoofed audio,'' in \emph{Interspeech 2024}, 2024, pp. 502--506.

\bibitem{EEND-EDA}
S.~Horiguchi, Y.~Fujita, S.~Watanabe, Y.~Xue, and K.~Nagamatsu, ``{End-to-End Speaker Diarization for an Unknown Number of Speakers with Encoder-Decoder Based Attractors},'' in \emph{Proc. Interspeech 2020}, 2020, pp. 269--273.

\bibitem{EEND-TA}
L.~Samarakoon, S.~J. Broughton, M.~Härkönen, and I.~Fung, ``Transformer attractors for robust and efficient end-to-end neural diarization,'' in \emph{2023 IEEE Automatic Speech Recognition and Understanding Workshop (ASRU)}, 2023, pp. 1--8.

\bibitem{AED-EEND}
Z.~Chen, B.~Han, S.~Wang, and Y.~Qian, ``{Attention-based Encoder-Decoder Network for End-to-End Neural Speaker Diarization with Target Speaker Attractor},'' in \emph{Proc. INTERSPEECH 2023}, 2023, pp. 3552--3556.

\bibitem{AED-EEND-EE}
S.~W. Zhengyang~Chen, Bing~Han and Y.~Qian, ``Attention-based encoder-decoder end-to-end neural diarization with embedding enhancer,'' \emph{IEEE/ACM Transactions on Audio, Speech, and Language Processing}, vol.~32, pp. 1636--1649, 2024.

\bibitem{EEND-NAA}
M.~Rybicka, J.~Villalba, N.~Dehak, and K.~Kowalczyk, ``{End-to-End Neural Speaker Diarization with an Iterative Refinement of Non-Autoregressive Attention-based Attractors},'' in \emph{Proc. Interspeech 2022}, 2022, pp. 5090--5094.

\bibitem{EEND-GLA}
S.~Horiguchi, S.~Watanabe, P.~García, Y.~Xue, Y.~Takashima, and Y.~Kawaguchi, ``Towards neural diarization for unlimited numbers of speakers using global and local attractors,'' in \emph{2021 IEEE Automatic Speech Recognition and Understanding Workshop (ASRU)}, 2021, pp. 98--105.

\bibitem{baevski2020wav2vec}
A.~Baevski, Y.~Zhou, A.~Mohamed, and M.~Auli, ``wav2vec 2.0: A framework for self-supervised learning of speech representations,'' \emph{Advances in neural information processing systems}, vol.~33, pp. 12\,449--12\,460, 2020.

\bibitem{liu2021pay}
H.~Liu, Z.~Dai, D.~So, and Q.~V. Le, ``Pay attention to mlps,'' \emph{Advances in neural information processing systems}, vol.~34, pp. 9204--9215, 2021.

\bibitem{hendrycks2016gaussian}
D.~Hendrycks and K.~Gimpel, ``Gaussian error linear units (gelus),'' \emph{arXiv preprint arXiv:1606.08415}, 2016.

\bibitem{zhang2019p2sgrad}
X.~Zhang, R.~Zhao, J.~Yan, M.~Gao, Y.~Qiao, X.~Wang, and H.~Li, ``P2sgrad: Refined gradients for optimizing deep face models,'' in \emph{Proceedings of the IEEE/CVF Conference on Computer Vision and Pattern Recognition}, 2019, pp. 9906--9914.

\end{thebibliography}

\end{document}